# Coherence manipulations of Poisson-distributed coherent photons for the second-order intensity correlation

Byoung S. Ham[1,2]
[1]Center for Photon Information, School of Electrical Engineering and Computer Science, Gwangju Institute of Science and Technology, 123 Chumdangwagi-ro, Buk-gu, Gwangju 61005, South Korea
[2]Qu-Lidar, 123 Chumdangwagi-ro, Buk-gu, Gwangju 61005, South Korea
(September 21, 2023; bham@gist.ac.kr)

**Abstract**
Unlike one-photon (first order) intensity correlation, two-photon (second order) intensity correlation is known to be impossible to achieve by any classical means. Over the last several decades, such quantum features have been intensively demonstrated for anti-correlation in the Hong-Ou-Mandel effects and nonlocal correlation in Bell inequality violation. Here, we present coherence manipulations of attenuated laser light to achieve such a quantum feature using pure coherence optics. Unlike the common understanding of the two-photon intensity correlations, the present coherence approach gives an equivalent classical version to the known quantum approach. To excite the coherence quantum features between paired coherent photons, a selective measurement process plays an essential role in creating the inseparable joint phase relation between independent local parameters. The local randomness is also satisfied in both parties using orthonormal polarization bases of a single photon.

**Introduction**
According to the quantum theory [1-4], the first-order intensity correlation is equivalent to that of classical physics such as for a double-slit interference fringe. However, it is generally accepted that there is no equivalent classical version of the second-order intensity correlation to the quantum theory [4-9]. Over the last several decades, the Hong-Ou-Mandel (HOM) effects [10,11] and Bell-inequality violations [12,13] have been intensively investigated for the mysterious quantum features using entangled photons generated from the second-order $(\chi^2)$ nonlinear media [14]. The anti-correlation of the HOM effects on a nonpolarizing beam splitter (BS) between two entangled photons is known to have no equivalent classical version of destructive interference. Such a photon bunching phenomenon in the HOM effects has been used to test sub-Poisson distributed single photons [15], Bell measurement-based quantum teleportation [16,17], and entanglement swapping between quantum nodes [18,19]. On the contrary, a preliminary coherence approach has also been studied for the same anticorrelation of the HOM effect [20]. As a result, a $\pi/2$ phase shift between entangled photons has been analytically derived [20]. An experimental proof has been separately presented for a trapped ion system [21]. This fixed relative phase between paired photons does not violate the quantum mechanics of the uncertainty principle, as agreed on in the EPR discussions [22].

According to quantum mechanics, the phase of a single photon should be undetermined. Thus, the phase parameter of a photon has been discarded in the quantum approach for the quantum correlation [23-25]. However, this does not mean that an instantaneous phase of a single photon is not allowed at all. Instead, a statistical mean of the instantaneous phases is undetermined. More interestingly, defining a fixed relative phase between paired photons does not violate the uncertainty principle as in the case of the position and momentum relation of a pair of entangled photons propagating in opposite directions [22]. With no phase assignment to the photon pair, thus, the BS-resulting phase [20] applied to the quantum approach is a bit awkward. It is more natural to the coherence approach to understand the destructive interference via quantum superposition between paired photons. Even though coherence optics belongs to classical physics due to the Poisson statistics, a post-measurement can result in a mysterious quantum feature such as in a delayed-choice quantum eraser [27-29]. Here, we present a coherence analysis of the two-photon intensity correlation for both HOM effects and Bell inequality violations using coherence manipulations of Poisson-distributed coherent photons in a Mach-Zehnder interferometer (MZI). For this, the system is coherently treated.

**Results**

Figure 1 shows a schematic of the present coherence analysis for the second-order intensity correlation between doubly bunched photons to derive coherence solutions of the HOM effects and Bell inequality violations. The Poisson-distributed coherent photons are generated from an attenuated laser L, where doubly bunched photon pairs are probabilistically (randomly) generated by Poisson statistics at ~ 1% ratio to the single photons [30]. The laser L is vertically or horizontally polarized, where its spectral linewidth $(\delta f_L)$ is narrow enough to keep coherence concerning the delay lines of the MZI. The neutral density (ND) filters are used to adjust the mean photon number $\langle n \rangle \ll 1$ of the attenuated light. Using a 22.5°-rotated half-wave plate (HWP), random polarization bases of a single photon are provided in each MZI path. The polarizing beam splitter (PBS) predetermines the MZI photon characteristics distinguishable (particle nature), resulting in no interference fringes in the MZI output ports. For the coherently widened spectral bandwidth $\Delta$ of the paired photons (see the Inset), a synchronized pair of acousto-optic modulators (AOMs) is used in the MZI paths. The AOM's frequency scanning directions are opposite, resulting in a correlated frequency relation at $f_\pm = f_0 \pm \delta f_j$, where $f_0$ is the center frequency of the laser L. Here, the bandwidth $\Delta$ is assumed to be Gaussian distributed, satisfying $\Delta \gg \delta f_L$, which is inhomogeneous for the photon ensemble by definition. The path-length difference of the MZI is controlled by the delay lines $D_\pm$ beyond the AOM-induced ensemble coherence length. The piezo-electric transducer (PZT) is for the fine-tuning of the MZI path-length difference in the order of the wavelength of L.

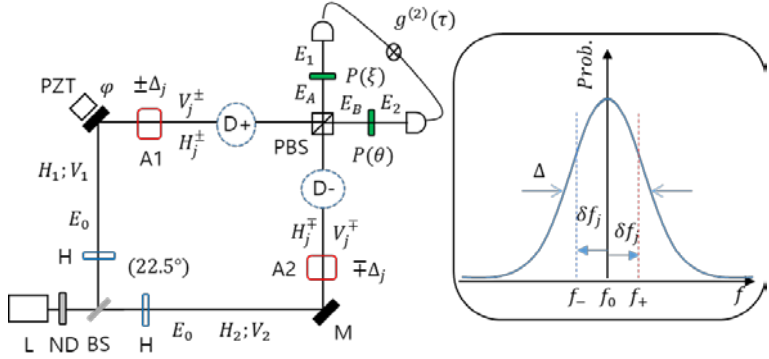

**Fig. 1.** Schematic of coherence manipulation for the quantum feature. A: acousto-optic modulator, L: vertically polarized laser, ND: neutral density filter, BS: nonpolarizing beam splitter, H: 22.5°-rotated half-wave plate, PBS: polarizing beam splitter, P: polarizer, PZT: piezo-electric transducer, D: delay line. Inset: AOM pair generated photon ensemble.

By ND-adjusted mean photon number $\langle n \rangle$ of coherent photons, the average time delay between consecutive single photons is set to be far longer than the coherence time $t_c$ of each photon, which is given by the inverse of $\delta f_L$, resulting in a statistical ensemble of photons for measurements. The MZI path lengths ($L_1$; $L_2$) are set to be far shorter than the coherence length $l_c$ of the laser L, satisfying a perfect coherence condition for all individual photon pairs. For the coincidence measurements, the time delay $\tau$ between any paired photons inside the MZI should reach far beyond the ensemble coherence time $\Delta^{-1}$. A polarizer P is inserted in each MZI output path for the quantum eraser. Each rotation angle of P represents an independent local parameter. For the full description of nonlocal correlation satisfying the Bell inequality violation, another identical MZI is needed [31]. For the cause-effect relation, the distance between BS and the polarizers must be longer than the ensemble coherence length given by $c\Delta^{-1}$, but far shorter than $l_c$. Thus, the violation of local realism is not for individual photon pairs but for the ensemble of them (see Discussion).

For the present coherence approach, the analysis for Fig. 1 follows pure coherence optics, satisfying the wave nature of a single photon in quantum mechanics. The conventional quantum approach is completely avoided. Thus, no quantized field or coherent state is involved. Using the matrix representation of the beam splitter (BS), coherence manipulations are applied to an arbitrary j$^{th}$ photon pair for HWPs, delay lines, AOMs, PZT, and polarizers. As a result, the following MZI output fields are obtained:

$$E_A^j = \frac{E_0}{2} e^{-i\delta f_j \tau} \left( \hat{H}^- - \hat{V}^+ e^{i(2\delta f_j \tau + \varphi)} \right), \quad (1)$$

$$E_B^j = \frac{iE_0}{2} e^{-i\delta f_j \tau} \left( \hat{V}^- + \hat{H}^+ e^{i(2\delta f_j \tau + \varphi)} \right), \quad (2)$$

where $\hat{H}^-$, $\hat{H}^+$, $\hat{V}^-$, $\hat{V}^+$ represent unit vectors of a single photon, whose superscript indicates the AOM-caused frequency detuning sign (see the Inset). Here, $E_0$ is the amplitude of a single photon. Due to the distinguishable photon characteristics in the quantum approach [32] or the Fresnel Arago law in coherence optics [33], the corresponding mean output intensities become uniform, $\langle I_A \rangle = \langle I_A \rangle = I_0/2$, where $I_0 = E_0 E_0^*$.

By the vector projections of the output photons onto the rotated polarizer, Eqs. (1) and (2) are rewritten as:

$$E_1^j = \hat{p}_\xi \frac{E_0}{2} e^{-i\delta f_j \tau} \left( \hat{H}^- \cos\xi - \hat{V}^+ \sin\xi e^{i(2\delta f_j \tau + \varphi)} \right), \quad (3)$$

$$E_2^j = \hat{p}_\theta \frac{iE_0}{2} e^{-i\delta f_j \tau} \left( \hat{V}^- \sin\theta + \hat{H}^+ \cos\theta e^{i(2\delta f_j \tau + \varphi)} \right), \quad (4)$$

where $\hat{p}_\xi$ and $\hat{p}_\theta$ are the polarizer's polarization axes whose angles are toward the counter-clockwise direction from the horizon axis [29]. Through the polarizers, the orthogonally polarized photon pairs become interfered. This is the so-called delayed-choice quantum eraser [13,29,34]. Here, the originally distinguishable photon characteristics become retrospectively converted into indistinguishable (wave nature) photons:

$$I_1^j(\tau) = \frac{I_0}{4} \left( \hat{H}^- \cos\xi - \hat{V}^+ \sin\xi e^{i(2\delta f_j \tau + \varphi)} \right) \left( \hat{H}^- \cos\xi - \hat{V}^+ \sin\xi e^{-i(2\delta f_j \tau + \varphi)} \right)$$
$$= \frac{I_0}{4} \left( \cos^2\xi + \sin^2\xi - 2\cos\xi \sin\xi \cos(2\delta f_j \tau + \varphi) \right)$$
$$= \frac{I_0}{4} \left( 1 - \sin 2\xi \cos(2\delta f_j \tau + \varphi) \right), \quad (5)$$

$$I_2^j(\tau) = \frac{I_0}{4} \left( \hat{V}^- \sin\theta + \hat{H}^+ \cos\theta e^{i(2\delta f_j \tau + \varphi)} \right) \left( \hat{V}^- \sin\theta + \hat{H}^+ \cos\theta e^{-i(2\delta f_j \tau + \varphi)} \right)$$
$$= \frac{I_0}{4} \left( \sin^2\theta + \cos^2\theta + 2\cos\theta \sin\theta \cos(2\delta f_j \tau + \varphi) \right)$$
$$= \frac{I_0}{4} \left( 1 + \sin 2\theta \cos(2\delta f_j \tau + \varphi) \right). \quad (6)$$

In Eqs. (5) and (6), the physical distance between BS and the polarizers must be beyond the light cone to violate the cause-effect relation [35]. This violation is not for individual photon pairs but for the ensemble photons, because the information is meaningful with an ensemble of bandwidth-distributed individual photons, as intensively discussed for the superluminal lights in the early 2000s [36]. The delay term $\delta f_j \tau$ becomes effective at $\tau \neq 0$ (see Fig. 2). At $\tau = 0$, however, the corresponding mean values of Eqs. (5) and (6) are bandwidth independent:

$$\langle I_1(\tau = 0) \rangle = \frac{I_0}{4} (1 - \sin 2\xi \cos\varphi), \quad (7)$$

$$\langle I_2(\tau = 0) \rangle = \frac{I_0}{4} (1 + \sin 2\theta \cos\varphi). \quad (8)$$

For the diagonally rotated Ps at $\xi = \theta = \pi/4$, Eqs. (7) and (8) become $\langle I_1(0) \rangle = \frac{I_0}{4}(1 - \cos\varphi)$ and $\langle I_2(0) \rangle = \frac{I_0}{4}(1 + \cos\varphi)$. These are the coherence solutions of the quantum eraser for Fig. 1.

The selective measurements are conducted by a DC-cut AC-pass filter for the coincidence detection between two output photons. This is the coincident heterodyne detection for the correlated intensity product:

$$R_{12}^j(\tau) = E_1^j(\tau) E_2^j(0)(cc)$$
$$= \frac{I_0^2}{16} \left( \hat{H}^- \cos\xi - \hat{V}^+ \sin\xi e^{i(2\delta f_j \tau + \varphi)} \right) \left( \hat{V}^- \sin\theta + \hat{H}^+ \cos\theta e^{i(2\delta f_j \tau + \varphi)} \right) (cc) A(\tau)$$
$$= \frac{I_0^2}{16} \left( \hat{H}^- \hat{H}^+ \cos\theta \cos\xi e^{i(2\delta f_j \tau + \varphi)} - \hat{V}^+ \hat{V}^- \sin\theta \sin\xi e^{i(2\delta f_j \tau + \varphi)} \right) (cc) A(\tau)$$
$$= \frac{I_0^2}{16} \left( \hat{H}^- \hat{H}^+ \cos\theta \cos\xi - \hat{V}^+ \hat{V}^- \sin\theta \sin\xi \right) (cc) A(\tau)$$
$$= \frac{I_0^2}{16} \cos^2(\xi + \theta) A(\tau), \quad (9)$$

where cc is the complex conjugate of $E_1^j$ and $E_2^j$. The term $A(\tau)$ is for $\tau$-dependent cross-correlation between Gaussian distributed paired photons. Thus, Eq. (9) represents the typical nonlocal correlation in a joint parameter relation [25,37]. The beauty of Eq. (9) is the self-cancellation of the phase term $2\delta f_j \tau + \varphi$, and thus the second-order intensity correlation can be represented by only local parameters of independent polarizers. The coherently derived joint parameter relation of two-photon correlation also represents the inseparable product basis in quantum correlation. Thus, Eq. (9) contradicts to our common beliefs that the second-order intensity (nonlocal) correlation cannot be obtained by any classical means. To interpret the coherently derived nonlocal quantum feature in Eq. (9), selective measurements play an essential role in exciting inseparable nonlocal correlation. Here, the nonlocal correlation originates in the violation of the cause-effect relation in the quantum eraser. Thus, nonlocal correlation satisfies the violation of local realism [31,35]. A full version of the nonlocal correlation is beyond the present scope and discussed elsewhere [31]. From this coherence analysis, it is clear that the entangled photon pair is a special case of two reduced product bases out of four tensor products.

**Numerical calculations**
  1. Classical correlation via direct intensity product

For $\xi = \theta = \pi/4$ and $\varphi = 0$, Eqs. (5) and (6) are rewritten as $I_1^j(\tau) = \frac{I_0}{4}(1 - \cos(2\delta f_j \tau))$ and $I_2^j(\tau) = \frac{I_0}{4}(1 + \cos(2\delta f_j \tau))$. These relations are the same as those for the BS-BS MZI system, except for the half-reduced intensity due to the polarizer. The numerical calculations of the intensity product $I_1^j I_2^j$ are shown in Fig. 2(a) as functions of $\tau$ and detuning $\delta f$ of the paired photons. For this, the coincidence detection is a prerequisite to separate the $j^{th}$ photon pair from others. Figure 2(b) is the sum of all $j^{th}$ photon pairs for $\tau$, showing the typical HOM dip, where the colored dots are for $\tau = 0, 2, 4, 6$ ($\sigma^{-1}$). The decay of the HOM dip is in the order of the ensemble coherence [10,11]. For $I_1^j I_2^j$, the $j^{th}$ Gaussian-distributed individual photon pair is weighted for $\delta f_j$, as shown in Fig. 2(c). Figure 2(d) is for individual cases of $I_1^j I_2^j(\tau)$, where the colored curves are from Fig. 2(a) for different $\tau$s. Each spectral sum for $\tau$ corresponds to each colored dot in Fig. 2(b). Thus, the nonexistence of (1,1) photon pair in both output ports has also been clearly understood upon the coherence approach via the coincidence detection.

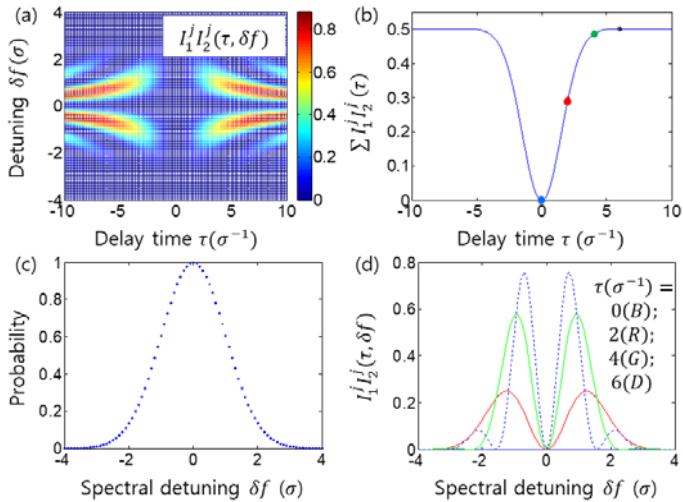

Fig. 2. Numerical calculation for intensity correlation based on Eqs. (5) and (6). (a) Individual photons (b) Sum of (a). (c) Gaussian curve. (d) Individuals of (b). $\sigma$ is the standard deviation in the Gaussian function $\exp\left(-\left(\frac{a\delta f - b}{c\sqrt{2}}\right)^2\right)$, where a=1, b=0, and c=5.

  2. Quantum correlation via heterodyne detection

From Eq. (9), $R_{12}^j(0) = 0$ result in for $\xi = \theta = \pi/4$. This is the direct result of the anticorrelation of the HOM effects [10,20]. As $\tau$ increases, however, $R_{12}^j(\tau) > 0$ results in due to the walk-off of Gaussian functions between paired photons. At $\tau \gg t_c$, $R_{12}^j(\tau \gg t_c)$ converges to the classical lower bound of $g^{(2)} = 0.5$, indicating completely incoherent and individual particles [12]. This is the correct definition of 'classical' in conventional quantum approach. In other words, classical means incoherent particles. The $\tau$-dependent cross-correlation between Gaussian pulses is also Gaussian, resulting in the same feature of Fig. 2(b). Thus, both classical $(I_1^j I_2^j)$ and coherently excited quantum $(R_{12}^j)$ show the same HOM effects. This is due to the single photon's self-interference in an interferometric system. Such a kind of HOM observations have already been conducted using independent light sources [38], where the same coherence analysis can be applied to the cavity optics of paired etalons.

**Discussion**
The coherence solution of the nonlocal correlation derived in Eq. (9) is due to the selective measurements at the cost of a 50 % measurement-event loss. The violation of local realism was conceptually understood with respect to the 'information' based on the ensemble bandwidth of photons. Even in this case, all measured individual photon pairs must be separately coherent for the interferometric system. Thus, the condition of the space-like separation is not for individual pairs but for the ensemble of them. Similarly, the violation of the cause-effect relation has no meaning to individual photon pairs, because information cannot be represented by a monochromatic light. This ensemble concept is also quite clear for the Schrodinger equation of a single particle. Thus, the causality violation derived in Eqs. (7) and (8) for the quantum erasers do not contradict the present coherence analysis. Instead, the quantum feature strongly requires coherence between paired photons. Without coherence, no quantum correlation occurs [39]. Regarding the HOM effects, both classical and quantum analyses matched each other upon coincidence detection. Thus, we can say that the HOM effects originate in the mean quantum superposition between paired single photon's self-interferences.

**Conclusion**
Using pure coherence optics belonging to classical physics, a coherence solution of the two-photon intensity correlation was analytically derived from Poisson-distributed coherent photon pairs for the nonlocal correlation violating local realism under local randomness. For the ensemble of coherent photon pairs obtained from coherence manipulations of an attenuated laser, the derived quantum feature originated in the selective measurements of the detected product bases, otherwise resulting in the lower bound of classical physics. Thus, the mysterious quantum feature was perfectly understood as a coherence feature via measurement-event modification. In that sense, the SPDC-generated entangled photon pairs are just preselected basis correlated photon pairs, as implied in the EPR discussion. The numerical calculations for the classical intensity product at coincidence detection showed the same anticorrelation of the HOM effects obtained from the coherence solution. This strongly supports the coherence feature of the HOM effects based on the ensemble coherence between paired photons' self-interferences. Both coherence solutions of the second-order intensity correlations of the HOM effects and nonlocal correlation satisfied the bandwidth-independent joint phase relation between independent local parameters, even though their origins are different. Thus, the conventional belief that there is no equivalent classical version of the second-order intensity correlation in quantum mechanics is now severely challenged.

**References**

1. Dirac, P. A. M. *Quantum mechanics*, 4th Ed. (Oxford University, 1958).
2. Mandel, L. & Wolf, E. Coherence properties of optical fields. Rev. Mod. Phys. **37**, 231-287 (1965).
3. Dimitrova, T. L. &Weis, A. The wave particle duality of light: a demonstration experiment. Am J. Phys. **76**, 137-142 (2008).
4. Feynman, R. P. QED: The strange theory of light and matter (Princeton University, 1985).

5. Rondon, R. Non-classical effects in the statistical properties of light. Rep. Prog. Phys. **43**, 913-949 (1980).
6. Glauber, R. J. The quantum theory of optical coherence. Phys. Rev. **130**, 2529-2539 (1963).
7. Ou, Z.-Y. Multiphoton quantum interference (Springer, 2007).
8. Scully, M. O. &Zubairy, M. S. Quantum optics (Cambridge University,1997).
9. Dirac, P. A. M. The principles of Quantum mechanics. 4th ed. (Oxford university press, London), Ch. 1, p. 9 (1958).
10. Hong, C. K., Ou, Z. Y. & Mandel, L. Measurement of subpicosecond time intervals between two photons by interference. Phys. Rev. Lett. **59**, 2044–2046 (1987).
11. Bouchard, F. et al. Two-photon interference: the Hong-Ou-Mandel effect. Rep. Prog. Phys. **84**, 012402 (2021).
12. Bell, J. On the Einstein Podolsky Rosen Paradox. Physics 1, 195-290 (1964).
13. Ma, X.-S., Kofler, J. and Zeilinger, A. Delayed-choice gedanken experiments and their realizations. Rev. Mod. Phys. **88**, 015005 (2016).
14. Zhang, C., Huang, Y.-F., Liu, B.-H., Li, C.-F., and Guo, G.-C. Spontaneous parametric down-conversion cources for multiphoton experiments. Adv. Quantum Tech. **4**, 2000132 (2021).
15. Shields, A. J. Semiconductor quantum light sources. Nature Photon. **1**, 215-223 (2007).
16. Bouwmeester, D., Pan, J.-W., Mattle, K., Eibi, M., Weinfurter, H. & Zeilinger, A. Experimental quantum teleportation. Nature **390**, 575-579 (1997).
17. Bayerbach, M. J., D'Aurelio, S. E. van Lookc, P. &Barz, S. Bell-state measurement exceeding 50% success probability with linear optics. Sci. Adv. **9**, eadf4080 (2023).
18. Pan, J.-W., Bouwmeester, D., Weinfurter, H. &Zeilinger, A. Experimental entanglement swapping: entangling photons that never interacted. Phys. Rev. Lett. **80**, 3891-3894 (1998).
19. Jin, R.-B., Takeoka, M., Takagi, U., Shimizu, R. &Sasaki, M. Highly efficient entanglement swapping and teleportation at telecom wavelength. Sci. Rep. **5**, 9333 (2015).
20. Ham, B. S. The origin of anticorrelation for photon bunching on a beam splitter. Sci. Rep. **10**, 7309 (2020).
21. Solano, E., Matos Filho, R. L. & Zagury, N. Deterministic Bell states and measurement of motional state of two trapped ions. Phys. Rev. A **59**, R2539–R2543 (1999).
22. Einstein, A., Podolsky, B. & Rosen, N. Can quantum-mechanical description of physical reality be considered complete? Phys. Rev. **47**, 777–780 (1935).
23. Gerry, C. C. &Knight,P. L. *Introductory quantum optics*. (Cambridge University, 2005).
24. Greenberger, D. M., Horne, M. & Zeilinger, A. Multiparticle interferometry and the superposition principle. Phys. Today **46** (8), 22-29 (1993).
25. Pan, J.-W., Chen, Z.-B., Lu, C.-Y., Weinfurter, H., Zeilinger, A. & Zukowski, M. Multiphoton entanglements and interferometry. Rev. Mod. Phys. **84**, 777-838 (2012).
26. Degiorgio, V. Phase shift between the transmitted and the reflected optical fields of a semireflecting lossless mirror is π/2. Am. J. Phys. **48**, 81–82 (1980).
27. Dimitrova, T. L. & Weis, A. Single photon quantum erasing: a demonstration experiment. Eur. J. Phys. **31**, 625 (2010).
28. Dou, L.-Y., Silberberg, Y. & Song, X.-B. Demonstration of complementarity between path information and interference with thermal light. Phys. Rev. A **99**, 013825 (2019).
29. Kim, S. & Ham, B. S. Observations of the delayed-choice quantum eraser using coherent photons. Sci. Rep. **13**, 9758 (2023).
30. Kim, S. & Ham, B. S. Revisiting self-interference in Young's double-slit experiments. Sci. Rep. **13**, 977 (2023).
31. Ham, B. S. Macroscopic quantum correlation using coherence manipulations of polarization-path correlations of a continuous-wave laser. arXiv:2308.04078 (2023).
32. Hardy, L. Source of photons with correlated polarizations and correlated directions. Phys. Lett. A **161**, 326-328 (1992).
33. Henry, M. Fresnel-Arago laws for interference in polarized light: A demonstration experiment. Am. J. Phys. **49**, 690-691 (1981).
34. Scully, M. O. and Drühl, K. Quantum eraser: A proposed photon correlation experiment concerning observation and "delayed choice" in quantum mechanics. Phys. Rev. A **25**, 2208-2213 (1982).
35. Kim, Y.-H., Yu, R., Kulik, S. P. and Shih, Y. Delayed "Choice" Quantum Eraser. Phys. Rev. Lett. **84**, 1-4 (2000).
36. Wang, L. J., Kuzmich, A. & Dogariu, A. Cain-assisted superluminal light propagation. Nature **406**, 277-

279 (2000).
37. Weihs, G., Jennewein, T., Simon, C., Weinfurter, H. & Zeilinger, A. Violation of Bell's inequality under strict Einstein locality. Phys. Rev. Lett. **81**, 5039-5042 (1998).
38. Deng, Y.-H. et al. Quantum interference between light sources separated by 150 million kilometers. Phys. Rev. Lett. **123**, 080401 (2019).
39. Herzog, T. J., Kwiat, P. G., Weinfurter, H. & Zeilinger, A. Complementarity and the quantum eraser. Phys. Rev. Lett. **75**, 3034–3037 (1995).

**Funding:** This research was supported by the MSIT (Ministry of Science and ICT), Korea, under the ITRC (Information Technology Research Center) support program (IITP 2023-2021-0-01810) supervised by the IITP (Institute for Information & Communications Technology Planning & Evaluation). BSH also acknowledges that this work was also supported by GIST GRI-2023.

**Author contribution:** BSH solely wrote the paper.

**Competing Interests:** The author declares no competing interest.
**Data Availability Statement:** All data generated or analysed during this study are included in this published article.